\documentclass[11pt]{article}
\usepackage[normalem]{ulem}
\pdfoutput=1 % if your are submitting a pdflatex (i.e. if you have

 \usepackage{jheppub}

\usepackage{url}
\usepackage{caption}
\usepackage{subcaption}
\usepackage{extarrows}

\usepackage[latin9]{inputenc}
\usepackage{float}
\usepackage{amsmath}
\usepackage{amssymb}
\usepackage{graphicx}
\usepackage{esint}
\usepackage{hyperref}
\usepackage{comment}
\usepackage{color}
\usepackage{microtype}
\usepackage{cleveref}
\usepackage{breakurl}
\usepackage{bbm}
\newcommand{\be}{\begin{equation}}
\newcommand{\ee}{\end{equation}}
\newcommand{\ben}{\begin{displaymath}}
\newcommand{\een}{\end{displaymath}}
\newcommand{\bea}{\begin{eqnarray}}
\newcommand{\eea}{\end{eqnarray}}

   \newcommand{\rf}[1]{(\ref{#1})}
\newcommand{\vp}{\varphi}
\def\E{{$E_{7(7)}$}}

\def\be{\begin{equation}}
\def\ee{\end{equation}}
\def\bea{\begin{eqnarray}}
\def\eea{\end{eqnarray}}
\def\ba{\begin{array}}
\def\ea{\end{array}}
\def\bit{\begin{itemize}}
\def\eit{\end{itemize}}

\def\na{\nabla}

\def\Tr{{\rm Tr}}

\def\a{\alpha}
\def\b{\beta}

\def\la{\lambda}

\def\vp{\varphi}

\def\na{\nabla}

\newcommand{\N}{\mathcal{N}}

\def\E{{$E_{7(7)}$}}

 \makeatletter

\allowdisplaybreaks

\makeatother

\makeatletter
\DeclareRobustCommand{\rcite}[1]{%
  \rcite@aux#1,\@nil{#1}%
}
\def\rcite@aux#1,#2\@nil#3{%
  \if\relax#2\relax
    Ref.~\cite{#3}%
  \else
    Refs.~\cite{#3}%
  \fi
}
\makeatother

\hypersetup{
    colorlinks = true,
    citecolor = {blue},
    linkcolor = {blue},
    urlcolor = {blue},
}

\def\id{\protect{{1 \kern-.28em {\rm l}}}}
\def\be{\begin{equation}}
\def\ee{\end{equation}}
\def\bea{\begin{eqnarray}}
\def\eea{\end{eqnarray}}


\def\Tr{{\rm Tr}}

\def \foot {\footnote}

\def \tr {{\rm tr}}
\def \ha {{1 \over 2}}

\def \ci{\cite}
\def \N {{\mathcal N}}

\def \E {{\mathcal  E}}

\def \d {\del}

\def\a{\alpha}
\def\b{\beta}

\def \del{\partial}
\def \a {\alpha}
\def \aa {{\rm a}}
\def\g{\gamma}

\def\ov{\over}
\def\E{{\mathcal E}}

\def\b{\beta}
\def\l{\lambda}

\def \k {\kappa}
\def\foot{\footnote}

\def \ci {\cite}

\def \d {\partial}
\def \Tr {{\rm Tr}}

\def \l  {\lambda}

\def \N {{\mathcal N}}

\def \N {{\mathcal N}}

\def \m {\mu}

\def \la {\label}

\def \l {\lambda}
\def\foot{\footnote}

\def \ov {\over}

\def\N{{\cal N}}

\def\cc{\circ}

\def \ha{{1\ov 2}}
\def \r {\rho}

\def \del {\partial}

\def \E {{\cal E}}

\def \la {\label}

\def \l {\lambda}
\def\foot{\footnote}

 \def \r {\rho}

\def \ov {\over}

\def \varpi {{\rm w}}
\def \OO {{\cal O}}

\def \ep {\epsilon}

\def\Tr{{\rm Tr}}
\def \d {\partial}

 \def \n {\nu}
  
\def \vp {\varphi}

\def \ed {\end{document}}
\def \te {\textstyle}

\def \ha {{{\textstyle{1 \ov2}}}}
\def \fo {{\textstyle{1 \ov4}}}

\def \d {\delta}

\def\Tr{{\rm Tr}}

\def\spa#1.#2{\left\langle#1\,#2\right\rangle}
\def\spb#1.#2{\left[#1\,#2\right]}

\def \ha {{\textstyle {1 \ov 2}}}

\def \vp {\varphi}

\def \nv { n_{\rm v}}
\def \ed {\end{document}} \def \iffa {\iffalse} \def \N  {{\cal N}} \def \vp {\varphi}
\def \l {\lambda} 
\def \ba {\begin{align}}
\def \ea {\end{align}}
  \def \r {\rho}

\def \la {\label}
\def \te {\textstyle} 
  
   \def \ha{{\textstyle {1\ov 2}}}

\def \nv {{\rm n_{\rm v}}}
\def \Nv {{\rm N_{\rm v}}}
\def \nt {{\rm n_{\rm T}}}
\def \Nt {{\rm N_{\rm T}}}

     \def \g {{\rm g}} 
\def \ed{\small
\bibliography{RefsCSG} \bibliographystyle{JHEP}  \end{document}
}
 \def \qq  {{ h}}
\def \rF {{\rm F}}
 \def \rT {{\rm T}}
 \def \rh {{\rm h}}
 \def \rq {{\rm q}}
 \def \pp {{p}}

   \def \rp {{\rm p}}\def \cc {{\rm c}}
\def \na {\nabla}
   \def \na {\nabla}
  \def \d {\delta} 
 \def \W  {{\cal W}}\def \cc {{\rm c}}
  \def \rp {{\rm p}}
   \def \g {\gamma}
\def \T  {{\cal T}}
 \def \rI  {{\rm I}}
 
\def \m  {\mu} \def \n  {\nu} \def \k {\kappa} \def \l {\lambda}

 \title{\rm { \LARGE \bf   On anomaly-free  4d  $\N=4$    and  6d   (2,0) 
 conformal  supergravities and UV finiteness of Poincar\'e   supergravities}}

\author[a]{R. Kallosh}
\author[,b]{and A.A. Tseytlin\footnote{Also at ITMP   and Lebedev  Institute.}}
\affiliation[a]{Leinweber Institute for Theoretical Physics at Stanford, 382 Via Pueblo, Stanford, CA 94305, USA}
\affiliation[b]{Abdus Salam Centre for Theoretical Physics, Imperial College London, SW7 2AZ, UK
}
\emailAdd{kallosh@stanford.edu}
\emailAdd{tseytlin@ic.ac.uk}

 \abstract{
We review   the   structure   of superconformal anomalies in 4d $\N=4$ conformal   supergravity (CSG)  coupled to  a number  $\Nv$  of $\N=4$ vector multiplets 
and 6d (2,0)  CSG  coupled to  $\Nt$ of  (2,0)  tensor multiplets.  Anomalies   cancel if $\Nv=4$ and $\Nt=26$ respectively. 
If the   CSG  part of the action is dropped and  $\Nv=6+ \nv$, the first theory is classically equivalent to the  4d $\N=4$ Poincar\'e supergravity (PSG)   coupled to $\nv$  vector multiplets, while the second one with  $\Nt=5+\nt$  is classically equivalent to the 6d (2,0) PSG coupled to $\nt$ tensor multiplets. 
We 
suggest
that these facts  imply  that divergences   in the  4d PSG   with  $\nv$ vectors  should be proportional to $\nv +2$
 and similarly  in the  6d PSG with $\nt$  tensors  to $\nt-21$. 
 This conjecture
  appears to be consistent with most of the 
   known   results of explicit scattering amplitude computations in these  4d and 6d  PSG theories,
  apart from one   coefficient   in  the 3-loop  mixed vector scattering  amplitude  computed in 
  arXiv:1305.4876. 
  }

\begin{document}

%  \hfill{\it Dedicated to the memory of  Igor Tyutin\ \ \ \ \ \  }

 % \vspace{2cm} 

  \hfill{\ \ \ \ \ \  Imperial-TP-AAT-01}

 \maketitle

 \setcounter{equation}{0} 
 \setcounter{footnote}{0}
 \setcounter{section}{0}

 \renewcommand{\theequation}{1.\arabic{equation}}
  \setcounter{equation}{0}
 
%{\small \tableofcontents
%}

% \tableofcontents{}

\parskip 3 pt

%\newpage

%%%%%%%%%%%%%%%%%%%%%%%%%
\section{Introduction}
\label{intro}
%%%%%%%%%%%%%%%%%%%%%%%%%
\renewcommand{\theequation}{1.\arabic{equation}}
\setcounter{equation}{0}

The study of UV divergences in supergravities is an old subject  %  (see, e.g.,  \ci{Bossard:2025ddf})
 that still brings  surprises. % \ci{Huang:2025nyr}.
It was   recently found  \ci{Huang:2025nyr}  that  the 1-loop 4-point   matter  scattering amplitude in 6d (2,0) Poincar\'e  supergravity (PSG)  coupled  to $\nt$  (2,0)  tensor multiplets  is   finite for $\nt=21$, despite the fact that it could  be expected to be  divergent for  any    $\nt\not=0$. 

The (2,0) PSG with 21 tensor multiplets   is, of course, known    to be special: 
 it is   gravitational  anomaly-free \ci{Townsend:1983xt}   and   also  
 %v3
 arises  from   compactification   
of type IIB supergravity  
%v2
or string theory on K3  \ci{Witten:1995em}.
However, the cancellation of the 1-loop UV divergences cannot be a direct consequence of the gravitational anomaly cancellation, as the chiral anomaly can only first show up in the finite part of the 1-loop effective action (this is in contrast to the scale anomaly, whose presence is visible already in the divergent part).

Here we  will  suggest  an explanation of the  result  of  \ci{Huang:2025nyr}
 % It just flips the {\1-loop finiteness of 4 tensors in the gravity sector} into  1-loop finiteness of 4 tensors in the matter sector, or $a$ in gravity  and $b$ in matter  with $a+b=4$. 
 %\draftnote{I now have doubts about this argument:  you used cancellation of anomaly at $\nt=21$ but 
  %1-loop  gravitational  chiral anomaly may    influence only higher-loop  divergences 
  %not 1-loop one. this is different from E8 story that was about 3-loop order. Thisis different from conformal anomaly that is correlated with divergences. }
   starting  with 
the fact that  the  (2,0)  PSG   with  $\nt$  tensor multiplets    admits a superconformal  formulation   \ci{Bergshoeff:1999db,Bergshoeff:1985mz} 
based on $\Nt=5+ \nt$   tensor multiplets  coupled to  6d   (2,0)  superconformal multiplet.\foot{A different  explanation 
of the result of  \ci{Huang:2025nyr} 
 based on the global  $SO(5,\nt)$  symmetry of this theory  (that   relates  1-loop finiteness  of  4-tensor  amplitude in 
  supergravity sector to 1-loop finiteness of 4-tensor  amplitudes   in the matter sector) 
  was  suggested in \ci{Kallosh:2025pmu}.}
In addition, we will use the observation  that   6d (2,0)  conformal supergravity (CSG)     coupled to $\Nt$  tensor multiplets
  is superconformal anomaly-free  for $\Nt=26$ \ci{Beccaria:2015uta,Beccaria:2015ypa,Beccaria:2017lcz}.

We will argue
 that   the superconformal  anomalies  of  (2,0) CSG   coupled to   $\Nt$  tensor multiplets  
  should   remain  the same   if one drops the  higher  derivative CSG part of the action, i.e. the anomalies of $\Nt$  tensor multiplets  coupled to (2,0)  superconformal multiplet  should also cancel for $\Nt=26$. 

   In this anomaly-free case, the  classical   equivalence between this superconformal model and  the 
   (2,0)  PSG   with  $\nt=\Nt-5=21$  tensor multiplets   should 
   %v3
   also   extend   to  the quantum level.\foot{%AT
   The cancellation of chiral gravitational anomaly in this 6d PSG  theory \ci{Townsend:1983xt}    then also may be viewed as a consequence of its connection to the anomaly-free superconformal theory.} 
   As the  absence of the superconformal anomalies implies,   in particular,   UV finiteness, this
    then leads to a prediction that the  (2,0) PSG with 21 tensor multiplets should be  finite. This  is 
    consistent  with the 1-loop  result of \ci{Huang:2025nyr},  and
    %v4
    we conjecture
     %  suggests
         that higher-loop amplitudes in this  6d  theory should also be finite, i.e.  in general should  scale as 
    $\nt-21$.

Remarkably, there is a close parallel between the  half-maximal supergravities in 4 and 6 dimensions. 
Indeed, similar observations  apply also to the  case of the 
 4d $\N=4$ Poincar\'e supergravity coupled to $\nv$ of   $\N=4$   vector multiplets.

At the classical level, this   theory   admits   a superconformal formulation
\cite{deRoo:1984zyh}   based on    $\Nv=6+\nv$   of   $\N=4$   vector 
multiplets  coupled to   the  4d $\N=4$ conformal supergravity  multiplet. 
The  4d  $\N=4$  CSG 
    can be made  superconformal  anomaly-free  by  adding  $\Nv=4$   vector multiplets 
     \ci{Fradkin:1983tg,Fradkin:1984pz,Fradkin:1985am,Romer:1985yg}.
As     the  superconformal  anomaly count  should  not change if one  omits    the  CSG  part of the total  action in the path integral (cf. \ci{Romer:1985yg,Carrasco:2013ypa}) this then   leads to  the conclusion 
 that  for $\Nv-4= \nv +2=0$  the 
4d $\N=4$ PSG   and its superconformal   version should be also quantum-equivalent  and thus both should   be    UV finite. 
  
 While the condition  $ \nv +2=0$ can  be    realised  only 
  in a non-unitary theory where 2 vector multiplets are  chosen to be ``ghost-like'' 
 (i.e. having opposite statistics   which  effectively reverses the sign of their  anomalies)
 the  above  argument implies  that   in general   the 
 divergences  in the $\N=4$ PSG with $\nv$   vector multiplets 
 should be   proportional to $\nv+2$. 
 This  indeed appears   to be consistent 
 with the   existing results  about the divergences  of the 
  1-loop \ci{Fischler:1979yk},  2-loop  and  3-loop \ci{Bern:2013qca}  and  also 4-loop \ci{Bern:2013uka}
  scattering amplitudes in this theory. 
  %v4
 One   exception is  the     $\zeta_3$ 
     coefficient   in  the 3-loop  mixed-multiplet  vector scattering  amplitude  computed in \ci{Bern:2013qca}
     that does not vanish for $\nv+2=0$   (see   discussion below). 
     This   may be indicating  some issue with our conjecture or that the computation in \ci{Bern:2013qca} is to be revisited.

We will start in section 2   with a review of anomalies in the 4d $\N=4$    conformal supergravity with $\Nv$ vector multiplets. 
Then in section 3  we   will   discuss the  relation  of this theory to the  $\N=4$  Poincar\'e   supergravity  with $\nv=\Nv-6$  vector  multiplets 
and the  resulting implications of  the superconformal anomaly cancellation for $\Nv=4$. 

Section 4 will be a summary of the anomaly cancellation  in the 6d (2,0)    conformal supergravity 
  coupled to $\Nt=26$   tensor multiplets. In section 5  we will  connect  this fact to the  prediction of  
   finiteness of  the 6d (2,0) Poincar\'e   supergravity  with $\nt=\Nt -5=21$   tensor  multiplets. 
  
  In Appendix A, we  will   collect   known results  about divergent  parts of scattering amplitudes in the 4d $\N=4$  PSG  with $\nv$   vector multiplets demonstrating that 
  %v4
  most of them  (apart from one coefficient  in one 3-loop amplitude)
   scale as $\nv+2$.    In Appendix B, we will  present   the   expression
    for the divergent  part  of  the 1-loop  scattering amplitudes in 6d (2,0)  PSG  with $\nt$ tensor multiplets.

%%%%%%%%%%%%%%%%%%%%%%%%%%%
\section{4d $\N=4$    conformal supergravity} %  with vector multiplets  } 
\renewcommand{\theequation}{2.\arabic{equation}}
\setcounter{equation}{0}

4d conformal supergravities   are     $\N \leq 4$ supersymmetric extensions  of the    $C_{\m\n\r\l}^2$  Weyl gravity   \ci{Kaku:1978pb,Bergshoeff:1980is}. 
Like   Weyl gravity, they are non-unitary   having higher derivative kinetic terms    but 
  are formally power-counting renormalizable  with  a single  coupling constant.
%but non-unitary  having higher derivative ghosts.
 % (ignoring topological term).
The corresponding 1-loop beta-functions   were  found  to be non-zero \ci{Fradkin:1982hx,Fradkin:1982uq} implying non-vanishing  
conformal anomaly for all $\N \leq 4$. 
As the Weyl symmetry here is gauged, this   implies quantum inconsistency \ci{Fradkin:1983tg}. 
For $\N =4$ CSG the  same   conclusion was reached also  from  the analysis  of  the chiral  $SU(4)$  R-symmetry gauge anomalies \ci{Romer:1985yg}, in agreement  
 with the fact that   all anomalies should  belong to the same $\N=4$ superconformal multiplet.
 % they   should be in the same  superconformal  multiplet with the Weyl anomaly. 

The 
   $\N=4$  CSG theory 
    can be made UV  finite  \ci{Fradkin:1983tg,Fradkin:1984pz,Fradkin:1985am}   and  also $SU(4)$    anomaly-free   \ci{Romer:1985yg}
 by coupling it    to  four    $\N=4$   abelian    vector  multiplets (or   to  $SU(2)\times U(1)$  $\N=4$  SYM).\foot{This was  shown   directly at  the 1-loop order but should be true to all  loop orders as the  beta-function 
   in $\N >1$   conformal supergravity   and the conformal anomaly of the  SYM theory     may  receive  contributions  only  from the first loop 
   (this  follows from formal   superspace  arguments as reviewed in  \ci{Fradkin:1985am}). 
   In the  case of $\N=4$ CSG  there  is also another reason  for the  1-loop exactness: 
  the  conformal  anomaly  is   tied by supersymmetry to  the  $SU(4)$   chiral anomaly  which 
   has 1-loop origin. Note also that the discussion of divergences/anomalies 
  %  We shall concentrate  on  terms  involving the Weyl tensor $C_{\m\n\k\l}$  and  $SU(4)$ gauge field  $\rF_{\m\n}$. 
    does not change if the   vector multiplets are replaced by  the $\N=4$ SYM theory with rank $\Nv$ gauge group.}

   In the  $SU(1,1)$ invariant   formulation of   $\N=4$  CSG    \ci{Bergshoeff:1980is} the 
   4-derivative complex   scalar $\vp$  did not couple  to  the Weyl graviton   and  the $SU(4)$    gauge field
   kinetic   terms. 
 It is    this  ``minimal"   $\N=4$ CSG  that  is ``induced"   (as the coefficient of the  log cutoff term) 
  in  the quantum  effective  action  of  $\N=4$ SYM   coupled  (in the standard $SU(1,1)$ covariant way  \ci{deRoo:1984zyh,deRoo:1985fq}) 
  to the    conformal supergravity background \ci{Liu:1998bu,Buchbinder:2012uh,Ciceri:2015nka}  or
  in the  classical action of   the 5d $\N=8$   gauged supergravity 
  evaluated  on the solution of the  AdS$_5$    Dirichlet  boundary problem \ci{Liu:1998bu}. 
 % However,  one  reason to expect  that there should  be another  inequivalent  version of $\N=4$ CSG with 
 %a  non-minimal coupling $f= e^{4 \vp} $  was  provided   \ci{6} by  dimensional reduction to 4d
 %  from 10d conformal supergravity    \ci{dw} (cf. eq.(4.23) in \ci{6}  and \rf{1e},\rf{3e}  below).
There  exists also a  ``non-minimal"   $\N=4$ CSG   with the action 
depending on  an  arbitrary   holomorphic function   $f$  of scalars
\ci{Butter:2016mtk,Butter:2019edc}. As was shown  in \ci{Tseytlin:2017qfd}
  the  divergences  of the CSG  theory  do not actually depend on a  particular  form of  the   function $f$, i.e. they 
  are the same   as in the ``minimal" theory, in   
  agreement  with the    chiral   anomaly count \ci{Romer:1985yg}.
  %  as required  by supersymmetry (the reason why that happens can be traced  back to the  holomorphicity of the  scalar   couplings  dictated by the  $\N=4$ supersymmetry). 
  %Thus  there is no  ``non-minimal"  $\N=4$ CSG  theory which is  UV  finite   by itself
 % but    as in the ``minimal" case   it 
  %can be made   anomaly-free and thus finite    by coupling it  to   four   $\N=4$  vector multiplets. 

        % Let us   briefly  review  some  basic relations. %  using  euclidean notation. 
   Let us review the anomalies of the   $\N=4$ CSG  coupled to $\Nv$   vector multiplets.
   %\foot{The discussion of divergences/anomalies 
  %  We shall concentrate  on  terms  involving the Weyl tensor $C_{\m\n\k\l}$  and  $SU(4)$ gauge field  $\rF_{\m\n}$. 
    %does not change if the   vector multiplets are replaced by $\N=4$ SYM theory with rank $\Nv$ gauge group.} 
 The corresponding  (Euclidean)  action   is  %(here we use Euclidean notation)
  %and  the scalar $\vp$   of $\N=4$   CSG. 
  %The Lagrangian of the $\N=4$  CSG  contains the following ``minimal" terms:
  ($i=1, .., 4; \ a= 1, ..., \Nv$)\foot{The CSG   multiplet  includes ($\del^4$) Weyl    vierbein $e^a_\m$,  $SU(4)$  ($\del^2)$ gauge
   field $V^i_{j \m}$  (with field strength $\rm F^i_{j\m\n}$) , 4  ($\del^3$) gravitini $\psi^i_\m$, 
    1 complex  ($\del^4$) scalar $\vp$, 
 4  ($\del^3$) spinors $\Lambda_i $,   10 ($\del^2$) complex scalars $E_{(ij)}$,  6  (anti)self-dual ($\del^2$)  tensors $T^{-}{}^{ij}_{\m\n}$  and 
 20 ($\del$) spinors $\chi_k^{[ij]}$.
 Note that the kinetic terms of the Weyl graviton  and the $SU(4)$ gauge field have opposite 
    signs in the    $\N=4$ CSG action. 
  This  is consistent   with  the fact   that  integrating out the  ``matter"  $\N=4$  vector multiplet  coupled to  conformal supergravity background   induces  the  $C^2$ term with positive (``asymptotically-free") sign   and the $\rF^2$ term with the 
  negative  (usual  ``non-asymptotically-free")  sign.}
%   and that provides a  possible   link  between   the  $F$  and $T $   
%contributions to the corresponding anomalies. } 
  \bea
  &\la{1} \te 
  S=  \int d^4 x \sqrt{g} \, L   \ , \qquad \qquad 
    L= \qq \,  L_{\rm CSG} +  \pp\,   L_{\N=4\,  \rm vec} \ , \ \ \ \\
& L_{\rm CSG} =   \fo   (C_{\m\n\k\l})^2 - \fo   \rF_{\m\n}^2    + ...\ ,\qquad   \qquad L_{\N=4\,  \rm vec}  =\fo  (F^a_{\m\n})^2  + ... \ , \la{2} 
 \eea
 % In what follows  we will suppress the internal  index $r=1, ...., 15$  on the $SU(4)$    field strength. 
 where $\qq$ and $\pp$ are the corresponding (inverse) coupling constants. 
  The    UV divergent part of the effective action   is  given by   %($\Lambda \to \infty$ is a cutoff)
    \be \la{3}\te
  \Gamma_\infty  = -  B_4 \log \Lambda \,   \ , \qquad   B_4= {1\ov (4\pi)^2}  \int d^4 x  \sqrt{g} \,  b_4 \ , \ \ \ \ \ \ \ 
   b_4 = { 2} \beta L_{\rm  CSG} \ , \ee
  where  the  beta-function coefficient    $\b$  is equal to -2   in the case of    pure  $\N=4$ CSG. 
  The conformal anomaly also depends on  the coefficient $a$  of the Euler number density:
% Representing the   conformal anomaly as  
    \be (4\pi)^2  \langle T^\m_\m\rangle = - a R^*R^* +  c  \big[C^2_{\m\n\k\l}  -  \rF_{\m\n}^2 + ...\big] = \b_1 R^*R^* +  \b   \big[  R^2_{\m\n} - \te {1\ov 3} R^2   -  \ha  \rF_{\m\n}^2 + ...\big] \ , \la{4}\ee
  %    , \ \ \ \   
   % W\equiv R^2_{\m\n} - {1\ov 3} R^2 = \ha (C^2_{\m\n\k\l} - R^*R^*), $   
   where 
   $\b_1=  c-a , \   \b=  2 c $. 
   One   finds  \ci{Fradkin:1982hx,Fradkin:1983tg,Fradkin:1985am}    that $\b_1=  c-a$  vanishes  separately for $\N=4$ SYM   and  $\N=4$   CSG  theories. 
    %which should  be a consequence of  their maximal $\N=4$ supersymmetry.
    
    The possibility of the  cancellation  of  the $\N=4$ CSG  beta-function  or $c$-anomaly 
     by coupling it to    $\N=4$  vector multiplets
     is a  consequence of the negative sign  of $c_{_{\rm \N=4\ CSG}}$ \ci{Fradkin:1982hx}:\foot{For a discussion of  this cancellation  from   the AdS$_5$ perspective \ci{Liu:1998bu} see section 5 in \ci{Beccaria:2014xda}.
    An   AdS$_5$ ``explanation"  of
      why the combination of the $\N=4$ CSG  and  four $\N=4$   vector multiplets is anomaly-free 
      %or why $c_{_{\rm \N=4\, CSG}} = (- 2) \times 2\,  c_{_{\rm \N=4\,  SYM} }$
involves their   indirect  connection 
 to the 5d  $\N=8$    supergravity: (i) the  partition functions   of   the 5d fields in  AdS$_5$
 and of  the corresponding 4d conformal fields  at the boundary 
  are  closely related % \ci{gk,bbt,16}
   (by the factor of -2)  so that  the  $a=c$  ``anomalies" of  the 5d   $\N=8$    supergravity and  of the  4d   $\N=4$   CSG 
 are also  related by the  factor of -2;  (ii) the 5d  $\N=8$   supergravity  may be viewed \ci{Gunaydin:1998uc} as  a  product of two   ``doubletons"  
 ($\N=4$   vector multiplets) and thus  their anomalies  are    related by the  factor  of  2.} 
    \be 
    c= c_{_{\rm \N=4\, CSG}} + \Nv\,  c_{_{\rm \N=4\,  vec} }= -1 + \Nv  \times \fo =\fo ( \Nv-4) \ .  \la{5}  \ee
   Since the $\N=4$ vector  multiplet   (or the SYM)    coupling is not renormalized,  the total theory is then finite if $\Nv=4$. 
%Similar  statement is true in 6 dimensions: conformal  a-anomaly of (2,0)   conformal supergravity is cancelled  by  coupling it to 26  (2,0)    tensor multiplets \ci{Beccaria:2015}, see below.
    
This   conclusion  is consistent  also  with the count of the non-abelian  $SU(4)$  gauge anomaly  \cite{Romer:1985yg}. Normalizing the
anomaly to the $d_{abc}$ symbols in the fundamental representation, %, $d^\rep{4}_{abc}$, 
a left-handed spinor $\psi^i$ % (or Majorana spinor whose left-handed part transforms in $\rep{4}$ of $SU(4)$)
 contributes $+1$;  the spinor  $\Lambda_i$ contributes $-1$;  the left-handed gravitino  $\psi^i_\m$ contributes  
 $+4$; $T^{ij}_{\m\n}$ does not contribute (the antisymmetric representation of $SU(4)$  is  real); 
$\chi^{[ij]}_k$  contribution is -7. Thus 
% \foot{One may use    that for $SU(N)$  
%   case  $d^{{[ij]}_k}_{abc}=\frac{1}{2}(N^2-7N-2)d_{abc}^\rep{N}\big|_{N=4} = -7 d_{abc}^\rep{4}$.} 
  the  total  $SU(4)$  gauge anomaly is 
\be
A^{\rm SU(4)}_{\rm CSG}= A_{\Lambda}   +  A_{\chi}  +  A_{\psi_\m}  =
  - A_{1/2}  -7A_{1/2}  + 4  A_{1/2} =  -4 A_{1/2} \ , \qquad
  A_{1/2}\equiv  { 1 \ov 24 (4\pi)^2}  \Tr ( \rF\rF^*) \ . \la{9}
 \ee
Adding   $\Nv$    vector multiplets    with the left-handed spinor $\psi^i$     in the fundamental  representation 
of $SU(4)$   we get 
\be 
%v3
A^{\rm SU(4)}_{\rm VM}= \Nv  A_{1/2}  \ , \ \ \ \ \ \ \  \ \ \ \ \ \ \ \ 
A^{\rm SU(4)}_{\rm CSG} + A^{\rm SU(4)}_{\rm VM}= (\Nv-4)  A_{1/2}\ ,  \la{10} 
\ee
i.e.  the gauge anomaly also cancels for $\Nv=4$. 

%\newpage
  
  \section{4d $\N=4$  Poincar\'e supergravity } % with vectors multiplets}
  \renewcommand{\theequation}{3.\arabic{equation}}
\setcounter{equation}{0}

The   classical 
${\cal N}=4$ Poincar\'e supergravity  admits   a superconformal formulation
\cite{deRoo:1984zyh}   based on    {six }   $\N=4$   vector 
multiplets  coupled to  fields of  the $\N=4$ conformal supergravity  multiplet.\foot{This is a  generalization of the familiar classical equivalence between 
the   %AT
``wrong-sign'' 
   conformal scalar theory %(here we assume Minkowski signature)
  $L= \del_\m \phi \del^\m \phi +  {1\ov 6} R \phi^2  $  
and the Einstein  theory $L= \kappa^{-2}R$. It holds in the  vacuum where the Weyl symmetry is spontaneously broken, $\langle \phi ^2\rangle = 6\kappa^{-2}$ (for a review  see, e.g.,
%v2
 \ci{Deser:1970hs,Freedman:2012zz}).
This does not imply a priori a direct relation between the two theories at the quantum level,  unless (super)conformal anomalies
 cancel (cf.  \ci{Kallosh:1974jy,Deser:1975sx,Ferrara:2012ui}).}

To get just the standard $\N=4$ PSG  action,   the 
  higher-derivative action of the pure  $\N=4$    CSG    should   not    be  
  added, i.e.   one  should   set  $\qq=0$ in \rf{1}.
 The  six   vector fields  couple  via their field strengths  $F$   to the six (anti)self-dual  tensors  $T_{\m\n}$ 
of the CSG multiplet   and after gauge fixing the  conformal   symmetry and S-supersymmetry 
and elimination of  other  auxiliary fields  these  produce  the six  physical  vector  gauge fields  of the PSG theory.\foot{To get the   physical sign of  the PSG  action one is to  add  
 the  6  vector  multiplets with the  negative sign of  their kinetic term  (i.e. with the corresponding $\pp$  coefficients 
  in \rf{1}   being   negative). 
 As the  fields of these 6   multiplets  play only   an auxiliary (Stueckelberg) role, 
 that does not  spoil the  unitarity of the resulting $\N=4$ PSG  theory.}

  More generally, starting with the number 
    \be \Nv=6+ \nv  \la{6}  \ee
  of  $\N=4$ vector   multiplets coupled to the  background fields of the 
   $\N=4$ CSG (with 6  ``negative-sign''  vector  multiplets   and  $\nv$     ``positive-sign''    ones,   implying $SO(6,\nv)$ global symmetry) 
  we 
  end   up  with   the action  of $\N=4$ PSG  coupled to $\nv$  $\N=4$ vector multiplets  (that we  will denote as  PSG$_\nv$).

  Let us introduce the notation CVM$_\Nv$ for the   theory of $\Nv$   $\N=4$ vector multiplets coupled   to the fields 
  of $\N=4$  CSG,  assuming that  
   all the fields are dynamical, i.e. 
      integrated over   in the path integral (the corresponding action is thus   the  $\qq=0$  limit  of  \rf{1}). 
  While in  the case when $ \Nv=6+ \nv$   this  model  is  {classically} 
   related   (upon  partial gauge fixing) 
   to the PSG$_\nv$, 
    this   equivalence  will not survive  in general at the quantum level  due to  superconformal anomalies. 
 %  We will also use notation  PSG$_\nv$ for $\N=4$ Poincar\'e supergravity 
  %  coupled to $\nv$   $\N=4$ vector multiplets.
   
  Which are the superconformal anomalies in  the CVM$_\Nv$  theory?  
   As reviewed  in  section 2, in the theory  with the action 
   \rf{1}   that includes the  higher-derivative   CSG  term the anomalies  are  controlled by the coefficient $\Nv-4$  in \rf{5}.
   %What happens if we drop CSG term, i.e.   set $q=0$ in \rf{1}?  
   One   may   conjecture  that  (chiral) 
   anomalies  having  1-loop  nature 
     should not depend on  the coefficient $\qq$ of the CSG   action.  
     %AT
     However, 
       %but  this   ignores the fact that 
        to  define  anomalies requires  a regularization 
      based on   the spectrum of   relevant kinetic operators,  but  it may seem  that  just 
      dropping    the  CSG   action  will  leave  the CSG    fields without kinetic terms.
      \iffa    However, 
   %  this is  not  so as  
     they mix with the   fields of the  
     %there  are  additional couplings  to the 
     vector multiplets  as in \rf{1} that do have standard kinetic terms.
     %and as a result may effectively contribute to the anomalies.   %:    this   leads to  a mixing  between the  fields of the  two multiplets.  
     \fi 
     
    In  fact, it is natural to expect  that      anomalies of the CVM$_\Nv$     should be   the same as in the 
    CSG + CVM$_\Nv$, i.e.  they should not depend on $\qq$ in \rf{1} as long as $\Nv$ is non-zero
    %AT
    (cf. footnote 14  in  \ci{Carrasco:2013ypa}). 
    %   .\foot{An argument in favour of this was  given in  \ci{Carrasco:2013ypa} (see  footnote 14 there).} 
     Adding   the CSG action   to the  vector multiplet  action should 
 not change  chiral anomalies as this is like adding a  higher-derivative regulator.
 %  (e.g.     for a  chiral fermion   $ D \to   D   +    D^3 $).
  %\foot{Also,  as  the  chiral gauge anomalies  may appear only at  1-loop order  they  
  %cannot depend   on the  coupling  constant 
% $\qq$ in  the   combined  Lagrangian in \rf{1}.}
 % i.e.    $ L=  \qq  ( C^2_{\m\n\k\l} +  ...) +  p (F^2_{\m\n} + ...)$.
   For  the $\Lambda_i$ and $\psi^i_\m$ fields  of the CSG multiplet 
     one gets  their kinetic operators 
 (in the scaling  symmetry broken phase) as 
 $\qq D^3 + p D$  and the  chiral  $SU(4)$  anomalies are  then the   same  for any value of the coefficient $\qq$
 (the $\qq D^2 + p$  factor      does not contribute  to the anomaly   \ci{Romer:1985yg}).
 %\foot{In general, anomalies 
%  do depend on which operator is used in regularization and this has 
  % to be the  kinetic operator in the action:  anomalies 
%    are related to  finite anomalous parts of  1-loop determinants. But what matters is  the chirality of the corresponding operator.}
%\foot{Fijikawa-type argument should always be  viewed from a physical 
% point of view  -- one cannot generate anomaly from non-derivative couplings}
%In general, one cannot generate an  anomaly from non-derivative couplings  %(modulo  definition of path integral measure)
%In general, one  may wonder how  some of the  CSG fields   may contribute to  the anomaly  when $\qq=0$  if   they  enter action only algebraically. 
The CSG fields that   for $\qq=0$  enter the   action  only algebraically  still mix 
 %The key point   that they 
 with the vector multiplet fields 
 that do have kinetic terms  and as a result may contribute to the anomaly.\foot{For $\qq=0$  the   antisymmetric tensor $T$   enters  the action only algebraically, 
   but it mixes with the vectors  of the  vector multiplets;  it should  then  be   defined 
 in terms of  derivatives of a complex vector  leading to a    non-trivial kinetic operator.
%  The six vector fields that  become  the six vector fields of the PSG
%multiplet  couple  via their field strengths  $F$   to the six (anti)self-dual rank-2 tensors  $T$ 
%appearing in the CSG multiplet.  That provides a     link  between   the  $F$  and $T $   
%contributions to the corresponding anomalies.
  Similar mechanism 
should  apply   in the case  of the   fermions $\chi$  that  do not 
 have  kinetic term  for  $\qq=0$ 
 but   mix  \ci{deRoo:1984zyh}  with the  fermions of  the vector multiplets.} 
 %That may be similar to mixing between 
 %The relation between anomalies of the CSG +VM and PSG systems 
% is thus not  completely 
% straightforward as  we need to  discard some contributions 
 %(of $\chi$ and  also of extra fermions of 6  vector  multiplets)  in the broken phase. 
 
While the  suggestion  that   the superconformal 
anomalies may be the same for $\qq=0$ and $\qq\not=0$  in \rf{1}    is thus supported by the analysis of the $SU(4)$   anomalies \ci{Romer:1985yg} 
it may seem    counter-intuitive  
   when applied to the   case of  the conformal   anomalies (related to  UV divergences) 
as they are  sensitive to the higher derivative terms in the kinetic operators.  The   agreement   between the  results 
  for 
 different types of  anomalies  which are parts of  the superconformal   multiplet  should  
 be a  non-trivial consequence of  the $\N=4$ supersymmetry.\foot{A  somewhat analogous 
  example  of an observable  % in     a  theory involving a combination of 
 %  2-derivative  and    4-derivative terms 
    that does not depend  on  the coefficient
 of the 4-derivative terms  is as follows. 
 Consider the 4d  gravity model  with %Einstein   Lagrangian   with extra   the  curvature-squared  terms 
  $L_{R + RR}=  R + \qq_1 R^2_{\m\n} + \qq_2 R^2$. 
 The tree-level S-matrix for  standard massless gravitons  in this theory is the same as 
 in the Einstein theory, i.e. it does not depend on  $\qq_1, \qq_2$  \ci{Dona:2015tra} (see also \ci{Beccaria:2016syk}). 
 There is a   simple  argument that  explains  why   that  happens. 
 The   generating functional for tree-level S-matrix is given by  the classical action  evaluated 
 on the solution of the classical field  equations  $g_{\m\n}=\eta_{\m\n} +  {\rh}^{(in)}_{\m\n} + O ((\rh^{(in)})^2)$. 
 Let us assume   that  $ {\rh}^{(in)}_{\m\n} $ represents  just  the massless   graviton. 
 As is well known, $R_{\m\n}=0$ solves also the  full field  equations  following from 
  $L_{R+ RR}$.  Then the   Einstein   theory tree-level S-matrix comes only 
   from the boundary  $\int 2K$ (GHY) term
  in the on-shell action   while the $R+ RR$ terms  vanish on the  $R_{\m\n}=0$  background, i.e.
   the  S-matrix does not depend on $\qq_1,\qq_2$.
  This argument  appears  to apply  also to   the  divergent part of the 1-loop 
   massless   graviton  sector  of the S-matrix:  as this theory is renormalizable,  the    divergent  part of the effective action   has the same form as the classical  action.
     %AT
   Similar   comments will  be valid  also  in the context  of the 6d (2,0) supergravity 
    where the $RR $ terms in the action are replaced  by $RRR$ ones in \rf{77} that also vanish on $R_{\m\n}=0$
    (cf.   also \ci{Maldacena:2011mk,chang2005}).}

   %Assuming  that this is the case, 
   Thus the anomalies  of  the CVM$_\Nv$  theory   should be controlled by the same coefficient  as in \rf{5},\rf{10}
   %, i.e.  by \foot{Once again, here $-4$  may be viewed as a  contribution of path integral measure of the  CSG fields   
   %while $\Nv$ is the contribution of the vector multiplets   coupled to  the  $\N=4$ Weyl multiplet fields.}  
   \be \Nv-4 = \nv  +2   \la{7} \ .  \ee
     This  anomaly    never cancels  for $\nv \geq 0$
   and thus  the CVM$_\Nv$   model    cannot be  quantum-equivalent  
   to the  unitary   PSG$_{\nv}$  theory. 
   
   One  may   formally  consider a non-unitary  model where we take   some  of  the $\nv$     vector multiplets in
   \rf{5},\rf{6}  to be 
    ``ghost-like'', i.e. with   anticommuting 
    %v3
    rather than commuting bosons and vice versa  for the fermions. This  will 
     reverse  the  signs of the corresponding  contributions to  the 1-loop effective action     and thus   also  the signs  of their  superconformal  anomalies, i.e.    will be  equivalent to  $\nv\to -\nv$.\foot{The  first 6   vector multiplets that  have negative  signs of their kinetic terms    will have  their path integral  contribution defined  using an analytic continuation 
    and  at the end  will be proportional to $\Nv=6$,   while   the contribution of the ``ghost''  (wrong statistics) multiplets 
     will be proportional to $-\nv$. }
Thus 
    if  we   admit  that  the number $\nv$
     can be effectively   negative if the   corresponding  multiplets are ``ghost-like'' 
       then the CVM$_\Nv$  model     will be  anomaly-free 
   if 
    \be \nv +2 =0 \la{8} \ .  \ee
 As a result,    CVM$_\Nv$   with $\Nv= 6 + \nv =4$    will   be    quantum-equivalent 
to the (non-unitary)  $\N=4$  PSG coupled to $\nv=-2$  (``ghost'') vector multiplets. 

The vanishing of superconformal anomalies  % in the theory with \rf{8} 
should imply, in particular, the  absence of  UV  infinities. 
We are then  led to the conclusion  
that   the S-matrix of the  $\N=4$  PSG  coupled to $\nv=-2$    vector multiplets    should  be {finite}. 
Since the  cancellation of anomalies   should be  1-loop exact, 
%v4
 we may then conjecture  %that   conclusion should 
that  this   should   then  be true  
%v3
 to all loop   orders.

%An implication  of the above   discussion   is 
%v4
An implication  of  this   conjecture  is 
that the  UV divergences in   the $\N=4$ PSG coupled to
any number  $\nv$  of $\N=4$   vector multiplets  
should   be in general  proportional to the  $\nv+2$ factor. 
Remarkably, this      appears to   be consistent with 
%v4
almost all of the known  results about  %what  is known  about  the
the   UV divergences in the 
scattering amplitudes  in this   theory  (see Appendix A). 

The 1-loop divergences in the $\N=4$ PSG$_{\nv}$ 
were  computed  in   \ci{Fischler:1979yk}  (see  also \ci{Fradkin:1983xs})  where 
   it was found that 
they are proportional to $(\nv+2) TT$   where here $T$   stands for  the stress tensor  of  vectors from  the  $\nv$ 
matter  multiplets.\foot{To recall, the 1-loop  divergences 
are absent  in the pure $\N$-extended  4d  PSG    for any $\N$.  Indeed, the    1-loop graviton amplitudes in 4d   are  finite not only in pure 
Einstein theory \ci{tHooft:1974toh}  % (thanks to $R^*R^*$  term being total   derivative) 
 but  also in the 
 Einstein theory coupled to  (scalar, vector, etc.) matter (cf.    \ci{Deser:1974cz}).  
 Thus  the  extension to supergravity is  just a consequence of supersymmetry. 
 In general, 
the divergent  part of the   1-loop effective action  has the structure $L_\infty \sim (RR + RT + TT) \log \Lambda $ 
which,  using field redefinitions or on-shell  relations,   can be  put into the  $TT$ form.  To find the  contribution to the S-matrix 
one is to evaluate $L_\infty$   on the ``in" solution of the classical equations of motion. If  one does  not consider 
external matter legs, i.e. sets  the corresponding ``in'' fields to zero  getting   $R_{\m\n}=0$, one   then  concludes that 
there are  no UV divergences
in the graviton  amplitudes. % (a remark to the contrary in \ci{Deser:1974cz}  was misleading). 
The  finiteness  of the  1-loop  4-graviton amplitudes in $\N=1,...,4$   4d supergravities 
was  verified also in \ci{Carrasco:2012ca}.}
%1209.2472 seems to suggest that N=4+n SGs have finite 4-graviton amplitudes also in d=6 at 1 loop.
%N=2 lifts to d=6, but the structure of the amplitude is different, and the arguments of 1209.2472 don't go through.
%Also, in d=8, even half-maximal SG develops a divergence.

The  4-graviton amplitudes are finite  also at 2-loop and  3-loop order. 
The   $\nv+2$    factor is found  again  in the divergent parts of the known  2-loop  and 3-loop  amplitudes  with
matter   vector  multiplet legs
\ci{Bern:2013qca}. 
  At 4 loops   there is a  divergence  in the 4-graviton  amplitude  \ci{Bern:2013uka}
  %\foot{The form of the divergence was 
   % argued \ci{Bern:2014lha} to be tied to the rigid 
% $U(1)$ duality anomaly of the theory (cf. \ci{Carrasco:2013ypa}), see also a discussion in appendix A.}
 and it is  again proportional to $\nv +2$.   
 
  %   and thus are finite in non-unitary theory satisfying \rf{8}.
    % This   gives    strong support to  the above  conjecture 
    %  that   PSG$_{\nv}$ theory  with  $\nv+2=0$ will   be finite   to all loop orders. 
      There   is, however, 
      %v4
         one   exception: there  
      %v3
      is   one  particular 
       3-loop amplitude (for scattering of vectors   from  two different multiplets)  computed in \ci{Bern:2013qca}
    that does not carry the overall   factor of  $\nv+2=0$  (cf. $\zeta_3$ term in   \rf{a8}). 
  This   may be an indication of  either some 
  %v4
  problem with    our argument  when applied to this case 
   or  that the computation of the $\zeta_3$  term in the 3-loop amplitude in 
   \ci{Bern:2013qca}  should be   revisited (see   comments   at the end of  Appendix A).

\newpage

%%%%%%%%%%%%%%%%%%%%%%%%%%%%%%%%%%%%%%%%%%%%%%%%
\section{6d   (2,0)  conformal supergravity} %   with   tensor multiplets}
\renewcommand{\theequation}{4.\arabic{equation}}
\setcounter{equation}{0}

%The  discussion in 6d  is closely analogous to  the one in 4d case.
%\ci{Bastianelli:2000hi,Henningson:1998gx,Butter:2016qkx,Casarin:2024qdn,Beccaria:2017dmw,Beccaria:2017lcz,Beccaria:2015ypa,Beccaria:2015uta}

The  6d  superconformal tensor calculus  was developed in 
 \cite{Bergshoeff:1986wc,Bergshoeff:1999db}  (for a  superspace formulation of $(2,0)$ conformal supergravity  and related references  see  also \cite{Kennedy:2025nzm}).
The 6d $(2,0)$ Weyl multiplet gauges the superconformal  group  $OSp(8^*|4)$  \cite{Bergshoeff:1999db}  and 
contains the  Weyl   graviton,   gravitini,  $USp(4)$ R-symmetry   gauge field $V^{ij}_\mu$   as well   as  the   antisymmetric tensor  $T^{ij}_{\mu\nu\lambda}$,  spinor $\chi^{ij}_k$   and  scalars $D^{ij, k\ell}$. 
The  symbolic  form of  the linearised  Lagrangian   is %dynamical 
  %field content 
 % of the  maximal (2,0)  6d conformal supergravity is  
       \be
\label{45}
L_{\text{(2,0)  CSG } } \sim  C_{\m\n\l\r} \del^2 C_{\m\n\l\r}   +
\psi^i _{\m\n} \del^3  \psi^i_{ \m\n} +  \rF_{\m\n}^{ij}  \del^2 \rF_{\m\n}^{ij}
+ T^{ij}_{\m\n\l} \del^4 T_{\m\n\l}^{ij} 
+ \chi^{ij}_k \del^3 \chi^{ij}_k  +  D^{ij,kl} \del^2  D^{ij,kl} ,
 %14\,\varphi+16\,\psi^{(3)}+ 5 T^{(4)}_{\m\n\l}  + 10\,V^{(4)}_\m+4\,\psi^{(5)}_{\mu}+e_{\mu}^{\ a}.
\ee
where $\psi_{\m\n}$ and $\rF_{\m\n}$  
   are the gravitino  and the vector field strengths. 
The 6d  $(2,0)$ tensor multiplet  includes 
 self-dual 2-form potential $B_{\m\n}$,  5  scalar fields $\phi$   and  4 fermions. 
Its coupling to the conformal supergravity multiplet  was discussed in 
% can be constructed by embedding the tensor multiplet supercurrent into the Weyl multiplet (see 
 \ci{Riccioni:1997np,VanHoof:1999xi,Bergshoeff:1999db}.
The non-linear   structure of the  6d (2,0) CSG action was  analysed   in \ci{Butter:2016qkx,Butter:2017jqu,Butter:2018wss,Casarin:2024qdn,Kennedy:2025nzm}.

%As in the 4d CSG case    the action of 6d CSG can be found  as an ``induced" one from the  1-loop  divergences of the  (2,0) 
%tensor multiplet in the CSG  background   or    from on-shell value of  action of extended supergravity in AdS$_7$ space
%evaluated on solution with Dirichlet  6d boundary conditions. 

Starting   with  a classically Weyl invariant 6d theory    one gets  the  1-loop   UV divergence  or  Weyl anomaly  which is the 
analog of  \rf{3}   with 4 independent coefficients  $\aa,\cc_1,\cc_{2},\cc_3$ 
 (for details see  \cite{Bonora:1985cq,Deser:1993yx,Bastianelli:2000hi,Boulanger:2007ab,Beccaria:2017lcz}; as in  4d  case  in \rf{4} 
 we omit covariant total derivative terms)
%the conformal anomaly of a classically  Weyl invariant theory  has the following 
%general  form 
%$  $   with  % \cite{Bonora:1985cq,Bastianelli:2000hi}
\ba
\label{41}  & \Gamma_\infty = - B_6 \log \Lambda \ , \ \ \ \ \ \ \ \ \qquad
 B_6 ={\te {1\ov (4 \pi)^3} }\int d^6 x \sqrt {g}\   b_6(x) \ ,\\  %\qquad \qquad 
 & b_6 = (4\pi)^{3} \langle T^\m_\m\rangle=  %= (4 \pi)^3  b_6= 
  -\text{a}\,\E_{6}+ \text{c}_{1}\,I_{1}+\text{c}_{2}\,I_{2}+\text{c}_{3}\,I_{3}\   .  \la{48}
%  &\la{411} E_{6} =- \ep_6 \ep_6 RRR, \quad 
%  I_1= C_{\alpha\mu\nu\beta}\,C^{\mu\rho\sigma\nu}\, C_{\rho}^{\ \alpha\beta}{}_{\sigma}, \ \  I_2 = C_{\alpha\beta}^{\mu\nu}\,C_{\mu\nu}^{\rho\sigma}\,
% C_{\rho\sigma}^{\alpha\beta}, \ \ \ I_3
 %= C_{\mu\alpha\beta\gamma}\, \nabla^{2} C^{\mu\alpha\beta\gamma} + ... \ 
\end{align}
Here $\E_{6} =- \ep_6 \ep_6 RRR$   is  proportional to the 6d Euler number density   and 
the   three  independent     Weyl invariants  $I_1,I_2,I_3$  are given by 
 \begin{align}
% E_{6} &= -\eps_{\mu_{1}\nu_{1}\mu_{2}\nu_{2}\mu_{3}\nu_{3}}
 %\eps^{\rho_{1}\sigma_{1}\rho_{2}\sigma_{2}\rho_{3}\sigma_{3}}\,
% R\indices{^{\mu_{1}\nu_{1}}_{\rho_{1}\sigma_{1}}}\,
% R\indices{^{\mu_{2}\nu_{2}}_{\rho_{2}\sigma_{2}}}\,
% R\indices{^{\mu_{3}\nu_{3}}_{\rho_{3}\sigma_{3}}}, \notag \\
 I_{1} &= C_{\alpha\mu\nu\beta}\,C^{\mu\rho\sigma\nu}\,C_{\rho}^{\ \alpha\beta}{}_{\sigma},
 \qquad I_{2} = C_{\alpha\beta}^{\ \ \mu\nu}\,C_{\mu\nu}^{\ \ \rho\sigma}\,
 C_{\rho\sigma}^{\ \ \alpha\beta}, \notag \\
 I_{3} &= C_{\mu\alpha\beta\gamma}\,(\nabla^{2}\,\delta^{\mu}_{\nu}+4\,R^{\mu}_{\nu}-
 \tfrac{6}{5}\,R\,\delta^{\mu}_{\nu})\,C^{\nu\alpha\beta\gamma}+\text{tot. deriv.}\ \la{42}
 \end{align}
   A  particular   combination of these  three  invariants  
 \be \la{55}
 \W_{6} =96 I_1 + 24 I_2 -  8 I_3   \ , \ee     has    special 
 properties:  (i) it vanishes  \ci{Bonora:1985cq}  on a Ricci flat background, and  (ii)  it admits a $(2,0)$  
locally   superconformal  extension \ci{Beccaria:2015uta,Butter:2017jqu,Butter:2016qkx}.

The strategy to  determine  the (2,0) CSG  action  may  be the same as in the 4d case  \ci{Liu:1998bu,Buchbinder:2012uh}: 
%%%%%%%%%%%%%%%%
\iffa 
   where  the 
action of $\N=4$  CSG  can be found as an { induced} one: 
  either as a local   UV singular part  of the  1-loop 
  effective action of $\N=4$  Maxwell  multiplet   
 coupled to $\N=4$ CSG  background 
   or as an IR  singular part of  the  value of the $\N=8, d=5$   gauged supergravity action evaluated 
 on  solution of the corresponding Dirichlet problem \ci{Liu:1998bu,Buchbinder:2012uh}. 
 Similarly, in the $d=6$  case we may  consider 
 \fi
 %%%%%%%%%%%%%%%%%%%%%%
 it can be  found either as a UV   divergent 
 part of  the induced  action for a 6d  $(2,0)$  tensor multiplet coupled to (2,0)  CSG  \ci{Bastianelli:2000hi}
  or as 
 an   IR singular part  \ci{Henningson:1998gx} 
 of the  action of  the maximal 7d gauged supergravity   \ci{Gunaydin:1984wc,vanNieuwenhuizen:1984iz}
 evaluated on the  classical solution with  the Dirichlet  boundary conditions in  AdS$_7$ vacuum.
%  This 
%definition of the CSG action   as a local part of   the induced  effective  action 
%guarantees the right symmetries and thus  allows in principle  to  determine    its full non-linear form.

%Related to (i)   and (ii)      is that  $\W_{6} $  
% appears, respectively,  as  the coefficient of the  logarithmic IR 
 %  divergence of the Einstein action in AdS$_7$   evaluated on the solution of Dirichlet problem \ci{Henningson:1998gx} and also 
%   as the  log UV divergence  of  the $(2,0)$  tensor multiplet   \cite{Bastianelli:2000hi}.
 The gravitational part of the 
  resulting (2,0) CSG  action  may be written  as\foot{The fact that  this action 
   is expressed in terms of the Ricci tensor  and is at most linear in the Weyl tensor
  implies  that it 
  can be rewritten  as a 2nd derivative action  involving  several tensors  of rank $\leq 2$; it 
    is uniquely selected by this requirement \ci{Metsaev:2010kp}.}
  \be
 \la{77}
S = \int  d^{6}x\,\sqrt{g}\, \W_6  \sim  \int d^{6}x\,\sqrt{g}\,\Big(R^{\mu\nu}\na^{2}R_{\mu\nu}
-\tfrac{3}{10}\,R\,\na^{2}\,R-2\,R^{\mu\nu\rho\sigma}\,R_{\nu\rho}\,R_{\mu\sigma}
-R^{\mu\nu}R_{\mu\nu}\,R+\tfrac{3}{25}\,R^{3}\Big).
\ee
In the presence of  at least  $(1,0)$  6d  supersymmetry  one expects that the   Weyl invariants $I_i$   are 
bosonic parts of only two possible 6d superinvariants, { i.e.}  the coefficients  $\cc_i$  in \rf{48}  should  satisfy  one linear relation. 
As  discussed in \cite{Beccaria:2015ypa},  the 
 free-theory  result  \cite{Bastianelli:2000hi}   and some 
holography-based arguments   \ci{Kulaxizi:2009pz} %, and studies in other contexts  
% \cite{Hofman:2008ar,Safdi:2012sn,Bueno:2015lza}
  indicate   that 
this  relation  is $\cc_{3} =\te - {1\ov 6} ( \cc_1 - 2 \cc_2)$.\foot{This  relation  indeed holds  not only 
 for the  divergence   of    the  free  $(2,0)$   
tensor multiplet   \cite{Bastianelli:2000hi}   but also   for 
the  conformal  anomaly  of the  large $N$ strong coupling limit of the  interacting  $(2,0)$  theory as described by  the 
supergravity in AdS$_7$  \ci{Henningson:1998gx}.} 

Thus in the (2,0)   case one should have in \rf{41}\foot{\la{f16}Note that on a Ricci flat  background one gets  then 
$
  b_{6}\big|_{R_{\m\n}=0}  = \,\big(\cc-\aa\big)\, \E_{6} \ , 
$
which is the analog of  the  familiar relation in 4d.}
\be  b_6 = - \aa\, \E_6 + \cc\, \W_6 \ ,    \qquad \ \ \  \ \ \  \cc= {1\ov 96}\cc_1 \ . \la{66}  \ee 
For a  single (2,0) tensor multiplet (denoted  below as   $\rT^{(2,0)}$)  one finds \cite{Bastianelli:2000hi}
\be \aa(\rT^{(2,0)}) = -{7\ov 1152} \ , \la{08}  \ \ \ 
\ \ \ \ \qquad \ \ \  \cc(\rT^{(2,0)})= - {1\ov 288} \ . \ee
The computation  of the  a-anomaly coefficient  for the (2,0) CSG multiplet  gives
  \ci{Beccaria:2015uta}\foot{An equivalent  computation  using a  relation to the   AdS$_7$   partition functions 
  was described in \ci{Beccaria:2014qea}. The  relevant field representation content is   determined, using, {e.g.},  the   relation to the 
  maximal   gauged 7d supergravity  (with   AdS$_7$  vacuum) which is   the bottom level of the Kaluza-Klein tower of multiplets corresponding to the   11d    supergravity compactified on $S^4$  
  (for a  translation  between fields of 
   7d  gauged and 6d conformal supergravities see  \ci{Bergshoeff:1999db,Nishimura:1999av}). }
%In this section, we discuss maximal gauged  7d supergravity and identify the induced 6d CFT 
%with  conformal supergravity \red{Liu-Ts, Bergshoeff, Sezgin, van Proeyen but they do not have action}.  
%The total  value of the a-anomaly  for the (2,0) CSG  multiplet  is found to be 
\be
\la{05}
\aa(\text{CSG}^{(2,0)}) = \frac{91}{576} = -26\,\aa(\rT^{(2,0)})\ . \ee 
For the CSG coupled to $\Nt$ tensor multiplets  we  then find 
 \be \ \aa(\text{CSG}^{(2,0)} + \Nt \rT^{(2,0)}) = (\Nt -26)\,  \aa(\rT^{(2,0)})\ ,  \la{04}
\ee
i.e. the   (2,0)  CSG      coupled to 26   tensor multiplets 
has    vanishing a-anomaly.

 One may also  compute the $\cc$-anomaly coefficient  in \rf{66} or the coefficient of the 
 Weyl-invariant $\W_6$ part of the 6d conformal anomaly  and thus the corresponding UV divergence.
 This was first done via  indirect methods  in \ci{Beccaria:2015ypa,Beccaria:2017lcz} and  was later  
   confirmed   in \ci{Casarin:2024qdn}.  As a result
  \be\la{445}   \cc(\text{CSG}^{(2,0)}) = \frac{13}{144} =  -26\,\cc(\rT^{(2,0)})  \ , \qquad \qquad 
 \cc(\text{CSG}^{(2,0)} + \Nt \rT^{(2,0)}) = (\Nt -26)\,  \cc(\rT^{(2,0)})\ . 
  \ee 
Thus   the system of (2,0)  CSG coupled to 26 tensor multiplets is superconformal 
 anomaly-free, i.e.   
 is  formally consistent   as a (non-unitary)  finite quantum field  theory.
 % MB-1 : rewritten similar to 6d
%$A_{4}\big|_{R_{\m\n}=0}  = (\cc-\aa)\,  E_{4}$ relation  in 4d  following from \rf{1}. 

Ref.  \ci{Beccaria:2015ypa} 
  proved that  (2,0) 6d conformal supergravity
coupled to  $\Nt=26$  tensor multiplets  is also   free  of  all chiral  anomalies. 
Since,  like the  4d  anomalies,   the 6d anomalies form a supersymmetry multiplet \ci{Howe:1983fr,Manvelyan:2000ef,Manvelyan:2003gc}
  one expects  to find  linear relations between their coefficients. 
   The  6d  chiral (R-symmetry and gravitational) 
  anomaly 8-form polynomial 
  %v3
   has the following general structure \ci{Frampton:1983ah,AlvarezGaume:1983ig,Zumino:1983rz} 
   \ba
&     \rI_{8} =  \frac{1}{4!}\big(\alpha\,\rq^{2}+\beta\,\rq\,\rp_{1}+
\gamma\,\rp_{1}^{2} +\delta\,\rp_{2} \big) \ ,\ \ \ \ \\
&
    \rq = \tr\,  \rF^2 \ , \qquad  \ \    {\rp}_1 = -\frac{1}{2}\,\tr\,  R^2, \qquad  \ \  \ {\rp}_2 = -\frac{1}{4}\,\tr\,   R^4 +\frac{1}{8}\,(\tr\,R^{2})^{2}\ ,  \la{145}
\end{align}
  where   $\vec \a = (\a,\b, \g, \d)$  
   are numerical coefficients.
   %\foot{The R-symmetry  gauge  bundle  forms 
 %  $c_1$ and $c_2$   should  not be confused with the   conformal  anomaly 
 %   coefficients $\cc_1,\cc_2$.} 
        The  fields contributing to  the gravitational and R-symmetry   anomalies in \rf{145}  are the 
       chiral   fermions,  conformal gravitini  and 
(anti)self-dual  rank 3  antisymmetric tensors. 
Their contributions  are summed up with  multiplicities corresponding to their 
$USp(4)= SO(5)$  R-symmetry  representations.
One finds that  all  4 chiral anomaly   coefficients 
%$\vec\alpha = (\a,\b,\g,\d)$  
are  proportional to the   anomaly coefficients  of  the $\rT^{(2,0)}$ 
multiplet\foot{This result is consistent with  the expected relation between the anomaly coefficients  $\cc -\aa= -  { 1 \ov 192} \delta $, cf. footnote \ref{f16}.}
\be\la{44} \vec\alpha \big(\text{CSG}^{(2,0)}\big) =-26\, \vec\alpha \big(\rT^{(2,0)}\big) \ , \qquad \qquad 
%\vec\alpha(\text{CSG}_{p}) = -2\big[6p(p-1)+1\big]\,\vec\alpha(\rT^{(2,0)})\ , \qquad \qquad 
\te \vec\alpha(\rT^{(2,0)}) = (1,\frac{1}{2},\frac{1}{8},-\frac{1}{2}) \ . 
\ee
Thus  the chiral anomalies of the (2,0) CSG  with $\Nt$  tensor multiplets  also  cancel if $\Nt=26$.

\section{6d  (2,0)  Poincar\'e  supergravity} %  with tensor multiplets}
\renewcommand{\theequation}{5.\arabic{equation}}
\setcounter{equation}{0}

Let  us recall that  starting with the 
 \be \Nt=5 + \nt \la{85}
 \ee  (2,0) tensor multiplets  coupled to   (2,0) CSG  background 
 and  spontaneously breaking %(or  {\em gauge-fixing})  
 the dilatation symmetry  by imposing a quadratic constraint on  $5\times (5+\nt)$ tensor 
 multiplet scalars   \ci{Bergshoeff:1999db,Bergshoeff:1985mz} 
 one   ends up with  a system of  $\nt$   tensor    multiplets coupled \cite{Romans:1986er,Riccioni:1997np} 
 to  the  (2,0)  6d  Poincar\'e supergravity  where 
 $5\nt$   scalars parametrize  the coset $SO(5,\nt) \ov SO(5) \times SO(\nt)$.
   
 In more detail, one imposes  the condition 
 $\eta^{IJ} \vp^{ij}_I \vp_{J kl} = M^4 \eta^{ij}_{kl}$,   where $\eta_{IJ}={\rm diag} (-...-+...+)$, $I,J=1, ...,5+\nt$, \  $i,j=1,...,4$ 
   and $M$ is a mass scale parameter that determines  the gravitational constant of the resulting 
 Poincar\'e supergravity.
 The  fields of  5 tensor multiplets      have negative signs of their  kinetic terms  to get 
 the standard  signs in  the Poincar\'e supergravity action. The  number 5 is  directly related to the 
 presence of 5    self-dual antisymmetric tensors $ T_{\mu\nu\rho}$  in the (2,0) CSG spectrum: 
 they couple to the antisymmetric tensor field strength $H_{\m\n\l}$ of the tensor multiplets 
 via  $ H^{\mu\nu\rho}  T_{\mu\nu\rho}$ term and (after the   conformal 
 and the S-supersymmetry  gauge  fixing)
  this  effectively  eliminates the  ``negative'' modes  of the 5 tensor multiplets. 
  This is a  6d  analog  of  the construction  of the 4d $\N=4$  Poincar\'e supergravity coupled to $\nv$ \     vector multiplets 
   by starting 
  with  $\Nv=6+\nv $   vector multiplets   in  $\N=4$ conformal supergravity background  discussed in section 3.
 % (there  the 6   vectors  couple to  6   tensors  $T_{\m\n}$  of $\N=4$   CSG via 
%  $ F^{\mu\nu}  T_{\mu\nu}$, { etc.}).  % (for  detailed discussions with  some applications see
 %  \cite{deRoo:1984gd,Ferrara:2012ui,Carrasco:2013ypa}).}
  
  If we   argue    as in the  4d case  that   dropping  the higher-derivative CSG term in the action 
    should  not change 
  the condition of  the superconformal anomaly cancellation,  then  the discussion in section 4   will imply    that 
  the   system 
   of $\Nt=5+\nt$  tensor  multiplets coupled to the  fields of (2,0) CSG  should   also be anomaly-free for $\Nt=26$:
    \be 
   \Nt =5 + \nt =26\ , \  \ \ \ \ \ {\rm i.e.} \ \ \ \ \ \  \ \  \nt=21 \ . \la{56} \ee
     This 
      leads to the conclusion  that this superconformal theory    should   be  quantum-equivalent 
   to the (2,0) PSG   coupled to $\nt=21$ tensor multiplets \ci{Beccaria:2015uta}.
   
     While in the 4d case  the  anomaly cancellation condition \rf{8} required  an unphysical    choice 
of the  number of vector multiplets, in the 6d case \rf{56}   it can be    satisfied  in a unitary theory
which was already  known  to be special --  
the one which is   gravitational  anomaly-free  and   originating also   from   compactification   
of type IIB supergravity on K3  \cite{Townsend:1983xt,Witten:1995em}.
In fact, the cancellation of   the gravitational anomaly in this  (2,0)  PSG theory  may  also be viewed as a consequence of its 
 relation to  the anomaly-free (2,0) superconformal theory.

As in the 4d case   in section 3   this then suggests  that  %the UV divergences  of the 
 since the superconformal version  of the (2,0) PSG    with  $\nt=21$ tensor multiplets  should be scale invariant, 
   % should   vanish   for  $\nt=21$, 
  the divergent parts of  the   scattering amplitudes  in the (2,0)  PSG theory  with $\nt$ tensors 
 should   have an overall  coefficient $\nt-21$  (which  is  the   counterpart of  $\nv+2$  in the 4d case). 

This conclusion  is  consistent with the result of the recent  computation of  
the UV divergent part of the  1-loop   S-matrix  in this 6d PSG   theory 
 \ci{Huang:2025nyr}  (see Appendix B). 
%It is  also   consistent with   an  argument  for  1-loop  finiteness  for $\nt=21$  based on  global symmetry considerations 
 %   \ci{Kallosh:2025pmu}.
     %(cf. also  \ci{Riccioni:2007hv,Heydeman:2018qqn}). 

Our conjecture about  quantum equivalence of the superconformal  formulation 
of the (2,0)  theory and its  (``spontaneously broken''  and partially gauge-fixed)  PSG version  
 suggests   that     higher-loop  divergences  
 in the (2,0) PSG with $\nt$   tensor multiplets  should   also scale as $\nt-21$. 
It would be  interesting to check this  prediction    by a   direct   computation of  the 2-loop S-matrix   in this theory.\foot{Let 
us note that the fact that the  theory of 
(2,0) PSG + 21 tensors  follows   in   a low-energy    approximation  from 
  type IIB   string theory on K3 \ci{Witten:1995em,Aspinwall:1996mn}  does not automatically imply that its  1-loop (and higher loop)  scattering amplitudes should be  UV  finite
  (though  this   fact  is  consistent   with  the gravitational chiral anomaly cancellation).  Indeed, 
  the  $\alpha'\to 0$  limit  of  string loop corrections  may be singular, reproducing potential  field-theory  UV divergences
  (cf.  \ci{Metsaev:1987ju,Minahan:1987ha}). For a discussion of  the  corresponding string scattering amplitudes 
  see   \ci{Lin:2015dsa,Berg:2016wux}.}

\

%%%%%%%%%%%%%%%%%%%%%%%%%%%%%%%%
\section*{Acknowledgements}
We  thank  {H. Johansson and }  R. Roiban  for  very  useful  discussions. The work of RK  is supported by  the Leinweber Institute for Theoretical Physics at Stanford and by NSF Grant PHY-2310429.
%v4
The work of AAT   is supported by the STFC grants ST/X000575/1.  % ST/T000791/1.

\newpage 
\appendix

\small 

\section{Divergences in 4d $\mathcal N=4$  PSG with $\nv $ vector multiplets}
\renewcommand{\theequation}{A.\arabic{equation}}
\setcounter{equation}{0}

This Appendix  summarises the explicit $\nv$ dependence of the divergent parts of the four-point  amplitudes  in 4d  $\mathcal N=4$ supergravity coupled to $\nv$ $\mathcal N=4$ vector multiplets 
 following \ci{Bern:2013qca,Bern:2013uka} (see also \ci{Bern:2012cd,Bern:2014lha}). 

 In     dimensional regularization  % one sets 
$
D = 4 - 2\epsilon.
$
The external states  are   labeled as:
 $H$  --  state in the $\mathcal N=4$ graviton multiplet;
 $V$ -- state in an $\mathcal N=4$ vector   multiplet; 
 $V^1, V^2$  -- states in distinct vector multiplets. % (requires $|\nv| \geq 2$).
$s,t,u$ are the  Mandelstam variables ($s+t+u=0$ for massless   kinematics). 
%$s = (k_1+k_2)^2, \  t = (k_2+k_3)^2, \  u = (k_1+k_3)^2$, with  $s+t+u=0$ for massless  kinematics.
The gravitational coupling is denoted by $\kappa$.
We  will  factor out the universal supersymmetric tree-level  contribution 
$
A_{\rm tree}\equiv A_0$, 
which encodes the external-state dependence and helicity structure. % (in the case of 16 supersymmetries). 

\iffa 
Following \cite{Bern:2013qca}, the matter dependence enters through the state-counting parameter
$
D_s = \nv + D = \nv + 4,
$  so that 
$D_s - 2 = \nv  + 2$.
\fi

{\bf One Loop}: \ \ 
\noindent All amplitudes with at least one external graviton multiplet state are 1-loop finite in $D=4$, 
while  the divergent  part of the 4-vector amplitude is proportional to $\nv +2$, i.e. 
\begin{equation}\la{a1}
\mathcal M^{(1)}_{ H_1, H_2, H_3, H_4}\Big|_{_{\text{div}}} = 0,
\qquad \ \ \mathcal M^{(1)}_{ H_1, H_2,V_3,V_4}\Big|_{_{\text{div}}} = 0 \ , 
\end{equation}
\begin{equation}\la{a2}
\mathcal M^{(1)}_{V_1,V_2 ,V_3,V_4}\Big|_{_{\text{div}}}
= -(\nv+2) \frac{3}{2(4\pi)^2\, \epsilon } \Big(\frac{\kappa^2}{2}\Big)^4 \ st \ A_0 \ .
\end{equation}
For  vectors from two different multiplets one  gets
\begin{equation}
\mathcal M^{(1)}_{(V^1)_1,(V^1)_2,(V^2)_3,(V^2)_4}\Big|_{_{\text{div}}}
= - (\nv+2)  \frac{1}{2(4\pi)^2\, \epsilon } \Big(\frac{\kappa^2}{2}\Big)^4 \ st \ A_0 \ . \la{a3}
\end{equation}

{\bf Two Loops}: \ \ \noindent Amplitudes with only graviton multiplet states  or only a single vector multiplet  state 
remain finite
% Divergences appear only in the mixed-flavor matter sector after subtraction of subdivergences.
%In particular, 
\begin{equation}\la{a4}
\mathcal M^{(2)}_{ H_1, H_2, H_3, H_4}\Big|_{_{\text{div}}} = 0
\ , \qquad \ \ \
\mathcal M^{(2)}_{H_1, H_2,V_3,V_4}\Big|_{_{\text{div}}} = 0
\ , \qquad \ \ \ 
\mathcal M^{(2)}_{V_1,V_2,V_3,V_4}\Big|_{_{\text{div}}} = 0.
\end{equation}
For states  from two different  vector multiplets   one finds  that   divergence is   again proportional to $\nv+2$
\begin{equation}\la{a5}
\mathcal M^{(2)}_{(V^1)_1,(V^1)_2,(V^2)_3,(V^2)_4}\Big|_{_{\text{div}}}
= -(\nv+2)^2  \frac{1}{4(4\pi)^4\, \epsilon^2} \Big(\frac{\kappa^2}{2}\Big)^6 \ s^2 t \ A_0 \ , 
\end{equation}
Note that \rf{a5} is consistent   with $M^{(2)}_{V_1,V_2,V_3,V_4}\Big|_{_{\text{div}}} = 0$ in \rf{a5}: 
  symmetrising  \rf{a5}   in external legs (and  using  that $st A_0$   is  invariant   under 
  this symmetrisation) one get a  factor of $s+t+u$ that  is equal to  zero. 
Thus  vanishing  of the divergent part of the  2-loop 
 4-vector  amplitude in \rf{a4} (as compared to  the  non-vanishing 1-loop  one  in \rf{a2}) 
  has a kinematical reason.\foot{\la{fr}We thank R. Roiban for pointing this out.}

{\bf Three Loops}: \ \ \noindent Amplitudes with external graviton multiplet states again remain finite
\begin{equation}\la{a6}
\mathcal M^{(3)}_{ H_1, H_2, H_3, H_4}\Big|_{_{\text{div}}} = 0
\ , \qquad \ \ \ \ \ \ 
\mathcal M^{(3)}_{ H_1, H_2,V_3,V_4}\Big|_{_{\text{div}}} = 0.
\end{equation}
For   all   vectors from the same matter  multiplet  one finds 
\begin{align}\la{a7}
& \mathcal M^{(3)}_{V_1,V_2,V_3,V_4}\Big|_{_{\text{div}}}
=  -\frac{1}{(4\pi)^6}  \,  \Big(\frac{\kappa^2}{2}\Big)^8\, B^{(3)}_\infty(s,t)  \ st \ A_0\ ,\\
%\frac{1}{(4\pi)^6} \Big(\frac{\kappa^2}{2}\Big)^8  \ st \ A_0\ ,
&  B^{(3)}_\infty(s,t)= 
(\nv+2)^2
\Big[
%v4:  1/2 -- > 1/4
\frac{1}{8\epsilon^3} (\nv+2)- \frac{1}{4\epsilon^2} + \frac{1}{4\epsilon}
\Big] \ (s^2 + t^2 + u^2) \la{a77}\ . 
\end{align}
If     vectors are  from two different   matter  multiplets  then 
\begin{equation}\la{a12}
\mathcal M^{(3)}_{(V^1)_1,(V^1)_2,(V^2)_3,(V^2)_4}\Big|_{_{\text{div}}}  
= -\frac{1}{(4\pi)^6}  \,  \Big(\frac{\kappa^2}{2}\Big)^8\, { \widehat B}^{(3)}_\infty(s,t)  \ st \ A_0\ , 
\end{equation}
\begin{equation}
{\widehat B}^{(3)}_\infty (s,t) = (\nv+2)^2 \Big[
\frac{1}{8\epsilon^3}  \big(\nv s^2 - 4 t u\big) +
\frac{1}{2\epsilon^2}   \ t u
 +
 \frac{1}{2\epsilon}  (s^2 + t u) \Big]    -    \frac{1}{\epsilon}    (7\nv -10)(s^2 + 2 t u)\,  \zeta_3
\ . \la{a8}
  \end{equation}
%%%%%%%%
The expression in \rf{a12},\rf{a8} is consistent   with \rf{a7},\rf{a77}  upon  symmetrising over the external  legs; in particular, the last  term 
in \rf{a8}  $\sim (s^2 + 2 t u)\,  \zeta_3$  symmetrizes to  zero (see footnote \ref{fr}).

{\bf Four Loops}:\ \ 
\noindent At four loops, only the divergence of the four-graviton amplitude is explicitly 
 known  at present (for any   value of $\nv$).  It is 
  proportional to a universal kinematic  tensor structure $\T$  \cite{Bern:2013uka} 
 built  out of   four  linearised Riemann tensors  %and  is encoding the kinematic dependence of the four-graviton counterterm
\begin{equation}
\mathcal M^{(4)}_{H_1,H_2,H_3,H_4}\Big|_{_{\text{div}}}
= \frac{1}{2304(4\pi)^8} \Big(\frac{\kappa^2}{2}\Big)^{10}\, C_\infty^{(4)}   \,  \T\ , \la{a9}
  \end{equation}
\be   C_\infty^{(4)} =   (\nv+2) \Big[ (\nv+2) 
   \big[
\frac{1}{\epsilon^2} \, 6\nv  + 
 \frac{1}{\epsilon} (3\nv+4) \big]  +   \frac{1}{\epsilon}  96(\nv-22)\,  \zeta_3 \Big] \ . \la{a16}
  \ee

{\bf Comments}:

The above expressions in \rf{a2},\rf{a3},\rf{a5},\rf{a77},\rf{a8},\rf{a16} are 
 proportional to $\nv+2$   in agreement with our 
 %v4
  conjecture 
     in section 3  that the $\N=4$ PSG  with $\nv+2=0$  should   be UV finite to all loop orders. 

%\bldraft{They have not computed the 4-matter UV divergences}
%  not sure what  you are   saying ?

The  only exception is the last $\zeta_3$ term  in \rf{a8}   that appears in the 3-loop   amplitude with 
  vectors from two different multiplets. 
  %v4
  One possibility  is that  for some reason our conjecture does not 
  apply to this case. It  is  also   possible  that there is 
 % It is   possible  that there is
   some   subtlety in how this term was  derived 
   in   \cite{Bern:2013qca}   and this  computation is  to be reconsidered.\foot{In the construction used in \cite{Bern:2013qca},   the $D_s=\nv + 4$ dependence originates from the non-supersymmetric Yang--Mills plus adjoint scalars factor obtained by dimensional reduction. At three loops  some contributions scale directly with the number of adjoint scalars and therefore with $D_s$ while other contributions arise from pure-gluon sectors or mixed gluon--scalar interactions, whose weights are not controlled by a single overall multiplicity factor.
The  $\zeta_3$   factor comes  from  genuine  3-loop massless master integrals. Different integral topologies enter the amplitude with different $D_s$-dependent numerator factors. After integration-by-parts  reduction, the $\zeta_3$ term  comes from a specific linear combination of these contributions and  superficially  there seems   to be  no apparent  reason for this combination to  scale as  $D_s - 2=\nv +2$.}
In particular, it may be that a  contribution of  some   evanescent \ci{Bern:2015xsa} 1-loop counterterm  is  missing 
(cf.  \ci{Bern:2017tuc,Bern:2017rjw}):  it 
 may    contain    an $\OO (\epsilon) \, \zeta_3$  part  from 1-loop diagram 
  that  may  lead to an extra ${1\ov \epsilon} \zeta_3$ term  after  being  multiplied  by $1\ov \epsilon^2$ in a
   3-loop contribution.\foot{We thank R. Roiban  for suggesting this possibility  and related discussions.}
   %v4 
   At the   same time, this appears to be somewhat unlikely as   low-loop subdivergences   are not generically  expected to produce 
    contributions with transcendental coefficients.\foot{We thank  the anonymous referee   for emphasizing this.}

Let us note  that $\nv+2$   appears also  to be 
  the coefficient  \ci{Carrasco:2013ypa}  of  the global $U(1)$  anomaly ($\del_\mu j^\mu \sim (\nv+2) RR^*$) 
 in the $\N=4$  PSG   with  $\nv$ vector  multiplets  (assuming particular $U(1)$  charge 
  assignments  appropriate to the PSG theory).  %(which are  different from the ones in superconformal  setting).  
  As a result, there are $U(1)$ violating  1-loop  graviton-graviton-scalar-scalar 
   amplitudes that have non-vanishing soft-scalar limits   \ci{Carrasco:2013ypa}.  
   %AT
   % (cf. \ci{Carrasco:2013ypa}).
   
   One  can 
    relate  the  presence of these $U(1)$ violating  amplitudes  to  particular terms in  divergent   parts of  higher-loop amplitudes.\foot{The form of the 4-loop divergence in \rf{a16}  was 
    argued \ci{Bern:2014lha} to be tied to this 
 $U(1)$ duality anomaly of this  theory. }
    As was pointed out in \ci{Bern:2017rjw,Bern:2019isl}  such anomalous 1-loop 
amplitudes  may  be   eliminated by adding  a finite local
counterterm   that  at the same time also cancels evanescent
contributions  relevant  for  the  analysis  of UV divergences in dimensional regularization. 
According to  \ci{Bern:2017rjw} 
``these
cancellations call for a reanalysis of the 4-loop ultraviolet
divergences previously found  in this theory  without the addition of
such counterterms''.  It may be  that the same   may  also apply  to the $\zeta_3$ term in the  3-loop 
divergence in \rf{a8}. 
    
 %   \bldraft{They never did anything since 2017 when they first promised to do it}
 % yes, Radu was saying that  Bern has a plan to revisit these computations ....

%\draftnote{We do not know about 4-loop UV divergences in 4v matter sector. If it is UV divergent at $n_v+2=0$, this is again a confirmation of another anomaly, which as far as I can see is the $SO(6, n_v)$ anomaly.}

%\newpage 
%%%%%%%%%%%%%%%%%%%%%%%%%%%%%%%%%%%%%%%%%
\section{Divergences in  6d  $(2,0)$  PSG with $\nt $ tensor multiplets}
\renewcommand{\theequation}{B.\arabic{equation}}
\setcounter{equation}{0}

   The  tree-level S-matrix for the  6d
(2,0)  Poincar\'e supergravity coupled to tensor multiplets  was found in 
% has an explicit compact formula (twistor/rational-map/CHY-like)
%v2
\ci{Heydeman:2018dje}. 
Recent developments in on-shell methods and the double-copy construction 
led to a closed-form expression  
 for the UV divergent part of the 1-loop   S-matrix  \ci{Huang:2025nyr}. 

 In dimensional  regularization  here  one sets 
  $D = 6 - 2\epsilon.$  %$s,t,u$ are  Mandelstam variables for massless  6d states. 
External states   are   labelled as:   $H$  --  state in the $(2,0)$  supergravity multiplet; 
 $T^f$ --  state in $f$-th  abelian $(2,0)$ tensor multiplet  ($f = 1,\dots,\nt )$.
 $A^{(1)}$ will denote  the {reduced} $1$-loop amplitude, with all universal supersymmetric and polarisation-dependent prefactors stripped off.

%Recent developments in on-shell methods and the double-copy construction have enabled the explicit computation of loop-level scattering amplitudes in this theory. In particular, closed-form expressions are now  available  for 

Let us recall first that 
while  the  pure Einstein  gravity in 4d is 1-loop  finite on-shell,  there   is a  non-trivial 1-loop   divergence
 remaining 
in the 6d  case 
\ci{VanNieuwenhuizen:1977ca,Critchley:1978kb,Bastianelli:2023oca}:
ignoring total  derivative terms  and  using the  on-shell  $R_{\m\n}=0$  condition 
 the  coefficient of  the 1-loop 
divergence  is proportional to 
$I_2 = (C_{\m\n\l\r})^3$ in \rf{42}, i.e.   in \rf{48}  one has  $b_{6}\big|_{R_{\m\n}=0} = - {9\ov 15120}  I_2$.

The   1-loop divergences  cancel out in the  pure   (2,0)  6d supergravity\foot{The theory  with $\nt\not=21$ 
 has chiral gravitational anomaly    and thus is formally inconsistent   beyond 1-loop  order but 
   one can still     analyse its 1-loop divergences as  the anomalous terms appear  only 
    in the finite  part of the 1-loop effective action.}
where  extended supersymmetry implies   that $b_{6}$ should be proportional to  the  invariant in \rf{55} 
which  vanishes  on $R_{\m\n}=0$  as follows from \rf{77}. Equivalently, 
 there is no   non-trivial $C^3$-like on-shell superinvariant in the case of  (2,0)  6d supersymmetry. % \ci{Kallosh:2025pmu}. 
The finiteness of the 1-loop  4-graviton amplitude in 6d  supergravity was   found  in \ci{Bern:2013qca}\foot{This reference 
explicitly discussed  (1,1) 6d supergravity  coupled to vector multiplets  but  one may 
  expect  that  the  conclusion about 1-loop finiteness of the 4-graviton amplitude should also apply  to the (2,0) theory (cf. also  \ci{Huang:2025nyr}).}
%\bldraft{[16] is only (1,1) 6d supergravity, it has only YM vector matter, no tensor matter. You can also  see it said explicitly  in  \ci{Bern:2012cd} in the beginning of sec. 2 they stress it is a non-chiral  (1,1) in contrast to a chiral (2,0) 
 %}
%\draftnote
%This implies  that there are no 1-loop   divergences  in  pure (2,0)   supergravity sector 
\begin{equation}\la{b4}
A^{(1)}_{H_1H_2H_3H_4}\Big|_{_{\text{div}}} =0 \ . 
\end{equation}
%%%%%%%%%%%%
The 1-loop amplitude with two external tensor multiplet states and two external graviton multiplet states was
also  found to be finite  \ci{Huang:2025nyr}
%\foot{It  was found  using the  double-copy and bootstrap methods. After tensor reduction, the would-be ultraviolet divergent contributions are proportional to chiral spinor structures that vanish identically in  6d   for any $\nt $.}
\begin{equation}\la{b44}
A^{(1)}_{T_1T_2H_3H_4}\Big|_{_{\text{div}}} =0 \ . 
\end{equation}
%%%%%%%%%%%%%%%%%%
\iffa 6d  $\mathcal N=(2,0)$ supergravity coupled to $\nt$ tensor multiplets is a chiral theory 
whose spectrum and interactions are strongly constrained by gravitational anomaly cancellation. The special value $\nt  = 21$ corresponds to the anomaly-free theory realized as the low-energy limit of type IIB string theory compactified on K3.
The full tree-level S-matrix for 6d
(2,0) supergravity coupled to 21 tensor multiplets has an explicit compact formula (twistor/rational-map/CHY-like)
\ci{Heydeman:2018qqn}. 
\fi
 %%%%%%%%%%%%%%%%%%
 The 1-loop amplitude with 4 external tensor multiplet states admits a flavour decomposition 
\begin{equation}\la{b1}
A^{(1)}_{T_1T_2T_3T_4}
= \delta^{f_1 f_2} \delta^{f_3 f_4}  F_s
+ \delta^{f_1 f_4} \delta^{f_2 f_3}  F_t
+ \delta^{f_1 f_3} \delta^{f_2 f_4}  F_u\ ,
  \end{equation}
  where the scalar functions $F_s$, $F_t$, and $F_u$ are related by crossing symmetry.
The reduced amplitude can be expressed in terms of the 6d scalar box integrals $B_{s,t}$ and $B_{s,u}$.
Its UV  divergent part   is  found to be \ci{Huang:2025nyr}  (cf. \rf{a2})\foot{The  finite  part  of $F_s$   may  be written as 
$
F_s\big|_{\text{fin}} 
= -\frac{1}{(4\pi)^3}
\big[
t^2 B_{s,t} + u^2 B_{s,u}
+ \frac{1}{12} ( \nt  - 21)\, s  \log(-\frac{s}{\mu^2})
  \big],
$
  with $F_t$ and $F_u$ obtained by  $s \leftrightarrow t$ and $s \leftrightarrow u$ \ci{Huang:2025nyr}.}
\begin{equation}\la{b3}
A^{(1)}_{{T_1T_2T_3T_4}}\Big|_{_{\text{div}}}
  = \,(\nt  - 21)  \frac{1}{8(4\pi)^3\, \epsilon } \Big(  s \, \delta^{f_1 f_2} \delta^{f_3 f_4} +   t\,  \delta^{f_1 f_4} \delta^{f_2 f_3}  + u\,  \delta^{f_1 f_3} \delta^{f_2 f_4}\Big)  \ . 
  \end{equation}
\iffa  \fi 
Thus  the  UV  divergence of the amplitude vanishes in 
 the  theory with $\nt  = 21$.
%which originates from anomaly-free superconformal  theory. 

%{\bf Comments}:

%At present, no explicit two-, three-, or 4-loop ultraviolet divergences have been computed for six-dimensional $\mathcal N=(2,0)$ supergravity with general $\nt $. While integrand constructions and double-copy representations exist for related half-maximal theories in various dimensions (cf. \ci{Johansson:2017srf}) , a closed-form extraction of multi-loop UV poles in $D=6$ as a function of $\nt $ has not yet appeared in the literature.
Let us  note again   that  this   cancellation of the 1-loop 
UV divergence  cannot be  a   direct consequence  of the fact that the 
 gravitational anomaly cancels \ci{Townsend:1983xt}  in  the 
(2,0) PSG with $\nt=21$: 
the  anomaly   shows up  only in the finite part of the 1-loop effective action.
%\foot{The argument for 1-loop 
%finiteness   \bldraft{ Yes, cancellation of the 1-loop 
%UV divergence it is not a consequence of absence of $SO(5)\times SO(21)$ anomalies, it is a consequence of $SO(5,21)$ symmetry of the action, as I said before.}

The explanation for \rf{b3} that we suggested in section 5 ties the  UV finiteness of the (2,0) PSG  with $\nt=21$ tensor  multiplets
  to the cancellation of   anomalies    in the superconformal  formulation of this theory   with 
  $\Nt=5+21$   tensor multiplets  coupled to the   (2,0)   conformal supergravity multiplet. 
  The conformal anomaly   cancellation     implies  in general the UV finiteness of  the  theory. 

A test of our conjecture would be   provided   by a 2-loop computation in this  (2,0) PSG theory. 
 %\bldraft{Below  you use the information about (1,1) supergravity, but remember [16] has no information about (2,0)}
 According to  \ci{Bern:2013qca}, in (1,1)  6d PSG  
 at {two loops } a divergence is  generically  expected  % in \bldraft{(1,1) supergravity} \ci{Bern:2013qca}
   already in the  gravitational sector  and the same may apply also  to the (2,0)  theory. 
The corresponding local counterterm  is of the  form
$
   \sim D^2 R^4 $.  It is   compatible with 16 supercharges 
   (which are % \bldraft{these 
   are different in (2,0) and (1,1)  cases).  
  % and has right dimension.
  The  expectation is that it should appear  again  with the  coefficient $\nt-21$.

  \

%\newpage 
\small
%\tiny
\bibliography{RefsCSGv2}
\bibliographystyle{JHEP-v2.9}

\end{document}